\documentstyle[aaspp4]{article}
\begin{document}
 
\title{{\ }
The effect of temperature anisotropy on observations of
Doppler dimming and pumping in the inner corona}
 
\vskip 0.2in
\author{\bf Xing Li$^{1,2}$,  Shadia Rifai Habbal$^1$, John Kohl$^1$
and Giancarlo Noci$^{3}$}
 
\vskip 0.05in
\affil{$^1$Harvard-Smithsonian Center for Astrophysics, Cambridge, MA 02138, USA}
\vskip 0.05in
\affil{$^2$Department of Earth and Space Sciences,
University of Science and Technology of China, \\Hefei, Anhui 230026, China}
\vskip 0.05in
\affil{$^3$Universita di Firenze, I-2025 Firenze, Italy}

\begin{abstract}
 
Recent observations of the spectral line profiles and intensity
ratio of the O VI 1032 {\AA} and 1037.6 {\AA} doublet
by the Ultraviolet Coronagraph
Spectrometer (UVCS) on the Solar and Heliospheric Observatory (SOHO),
made in coronal holes below 3.5 $R_s$,
provide evidence for Doppler dimming
of the O VI 1037.6 {\AA} line and pumping by the chromospheric
C II 1037.0182 {\AA} line. Evidence for
a significant kinetic temperature anisotropy of 
O$^{5+}$ ions was also derived from these observations.
We show in this Letter how the component of the kinetic temperature
in the direction perpendicular to the magnetic
field, for both isotropic and anisotropic temperature distributions,
affects both the amount of Doppler dimming and pumping.
Taking this component into account, we further show that the observation
that the O VI doublet intensity ratio is less than unity
can be accounted for only if
pumping by C II 1036.3367 {\AA} in addition to C II 1037.0182
{\AA} is in effect.
The inclusion of the C II 1036.3367 {\AA} pumping
implies that the speed of the O$^{5+}$ ions can reach
400 km/s around 3 $R_s$ which is significantly higher than the 
reported UVCS values for atomic hydrogen in polar coronal holes.
These results imply that oxygen ions flow much faster than protons
at that heliocentric distance.

\end{abstract}


\section{Introduction}

The Ultraviolet Coronagraph Spectrometer (UVCS) on SOHO has
proven to be a powerful tool for probing the physical conditions
in the inner corona (Kohl et al. 1995, 1997, 1998; Habbal et al. 1997; 
Noci et al. 1997). These ultraviolet measurements extend to
at least 3.5 $\rm R_s$ in polar coronal holes, and to 10 $\rm R_s$
in denser coronal plasmas such as streamers.
As described by Kohl \& Withbroe (1982), and Withbroe et al. (1982),
determinations of proton and heavy ion flow speeds in the inner corona
can be made using the Doppler dimming effect (see Hyder \& Lites 1970).
This method is applicable when ultraviolet coronal lines
are formed primarily by the resonance scattering of ions by the 
chromospheric or transition region radiation in the same
wavelength range, in addition to collisional excitation.
As ions flow outwards in the corona, 
the fraction of the spectral line formed by resonance scattering 
becomes Doppler-shifted out of resonance with the disk emission. 
Subsequently, only the collisional component then remains.
Of particular interest is the application of this technique to
doublets such as the O VI 1032 and 1037.6 {\AA} lines.
In a static corona,
the ratio of the line intensities 1032/1037.6 equals 4
when resonant scattering is dominant, 
and reduces to 2 when collisional excitation only is present.

Extending the concept of the Doppler dimming effect,
Noci et al. (1987) showed that determinations of flow speeds in the range
of 100 to 250 km/s could also be made.
With the assumption of isotropic kinetic temperatures, that study 
was based on the possibility of scattering of the C II 1037.0 {\AA}
line by the coronal $\rm O^{5+}$ ions 
when the ion flow speed is in the range of 100 to 250 km/s.
When this occurs, the ratio 1032/1037.6 of the two line intensities
is then expected to reach a minimum as the flow speed of the ions
increases as a function of heliocentric distance. 

Another important discovery from UVCS is the observation of
broader than expected profiles of
O VI 1032 and 1037.6 {\AA} lines in polar coronal holes. 
Since these profiles are measured
along the line of sight which is perpendicular to the radial direction
at the point of closest approach to the Sun, their widths yield a
measure of the $\rm O^{5+}$ kinetic temperature in the direction 
approximately perpendicular to the magnetic field.
UVCS observations also revealed that the O VI 1032/1037.6 intensity 
ratio reached a value of less than one at about 2.5 $R_s$ in coronal
holes. If the kinetic temperatures parallel to the magnetic field were
similar to the observed perpendicular temperatures, the O VI intensity
ratio could not reach a value near unity. 
Hence, observations of broad line profiles and a decrease in the
line ratios below a value of one
led Kohl et al. (1997) to conclude that a
significant kinetic temperature anisotropy exists
for coronal $\rm O^{5+}$ at a heliocentric distance of about 2.5 $R_s$.

Motivated by the UVCS observations, we illustrate in this Letter how the
component of the kinetic temperature along the line of sight,
for both isotropic and anisotropic temperature distributions,
affect the derivations of outflow velocities from the O VI intensity ratio.
However, by extending the effect of Doppler dimming and pumping
of the O VI 1037.6 {\AA} line to include
another almost equally strong chromospheric
C II 1036.3367 {\AA} line in the vicinity of the
C II 1037.0182 {\AA} line (Warren et al. 1997),
we show that the pumping by both C II lines
can account for the O VI line intensity ratios
recently observed by UVCS (Kohl et al. 1997).

\section{Effect of temperature anisotropy on Doppler dimming and pumping}

The resonant scattering component of a spectral line, $I_R(\nu)$,
integrated along a line of sight (LOS) chosen along the $x$-axis
can be written as (Kohl $\&$ Withbroe, 1982):
\[
I_R(\nu) = \frac{h B_{12}}{4 \pi \nu_0}
\int_{-\infty}^{\infty} n_i dx \int_\omega
[a+b({\bf \mbox{n} \cdot \mbox{n}'})^2] d \omega \times
\]
\[
 \int_{-\infty}^{\infty} I(\nu',\omega)d\nu'
\int_{-\infty}^{\infty}f(\mbox{\bf v})
\delta \left(\nu'-\nu_0-\frac{\nu_0}{c}
{\bf v \cdot n'}\right)
\]
\begin{equation}
\times \delta
\left( \nu_0-\nu+\frac{\nu_0}{c}{\bf v \cdot n}
\right) d{\bf v},
\label{eq:rad}
\end{equation}
where $B_{12}$ is the Einstein coefficient, $h$ is Planck's constant,
$I(\nu',\omega)$ is the
intensity of the chromospheric radiation at wave frequency $\nu'$.
The only photons that can be
scattered by an ion, O$^{5+}$ in the example considered here,
with a velocity ${\bf v}$, are those with $\nu' =
\nu_0 + (\nu_0/c){\bf v \cdot n'}$ where $\nu_0$
is the central wave frequency of the
oxygen spectral lines and ${\bf n'}$ is the vector
describing the direction of the incident chromospheric radiation.
$n_i$ is the number density of O$^{5+}$, ${\bf v \cdot n}$
is equal to $v_x$, $\omega$ is the angular direction, and $f({\bf v})$ is the
velocity distribution function describing the oxygen ions.
$[a+b({\bf \mbox{n} \cdot \mbox{n}'})^2]$ is the angular dependence of the
scattering process for the oxygen lines.
For O VI 1032 {\AA}, $a$ is equal to 7/8
and $b$ is equal to 3/8; for O VI 1037.6 {\AA}, $a$ is 1 and $b$ is 0 (Noci
et al. 1987).

If a bi-Maxwellian velocity distribution is assumed, the integral over
${\bf v}$ can be computed
analytically to yield an expression for $I_R(\nu)$ which can then
be integrated numerically over $x,~\omega,$ and $\nu'$.  
As an illustration, we define a rectangular coordinate system with
the +$x$-axis pointing towards the observer along the LOS (see Figure 1).
At any point along the LOS, the +$z$-axis is chosen to point towards
the Sun and to be perpendicular to the LOS
in the plane formed by the LOS and Sun center O.
The three angles, $\Phi$, $\theta$ and $\Psi$, defined in this
coordinate system are shown in Figure 1.

For a bi-Maxwellian velocity distribution, $f(\mbox{\bf v})$ can be written as
\[f(\mbox{\bf v})=\left( {m \over 2 \pi k T_ \parallel } \right)^{1/2}
\left( {m \over 2 \pi k T_\perp } \right) \mbox{exp}\left( {-m \over 2kT_\parallel} (v_\parallel-u)^2\right)
\]
\begin{equation}
\times \mbox{exp}\left( {-m \over 2kT_\perp} v^2_\perp \right) 
\end{equation}
where $u$ is the bulk outflow velocity along the radial direction, 
 $T_\parallel$ and $T_\perp$ are the kinetic temperatures
in the radial and perpendicular to radial directions with the 
assumption that the magnetic field expands radially. 

We define
\begin{equation}
\alpha_\parallel = {2kT_\parallel \over m} , ~ \alpha_\perp= 
{ 2kT_\perp \over m},
\end{equation} 
The last integration in (\ref{eq:rad}):
\[F=\int_{-\infty}^{\infty}f(\mbox{\bf v})
\delta \left(\nu'-\nu_0-\frac{\nu_0}{c}
{\bf v \cdot n'}\right)
\times ~~~~~
\]
\[
\delta
\left( \nu_0-\nu+\frac{\nu_0}{c}{\bf v \cdot n}
\right) d{\bf v}
\]
can then be transformed into the analytical expression:
\[F
= {A^{-1/2} \over \pi \sqrt{ \alpha_\parallel} \alpha_\perp} 
\mbox{exp}
 \left(-{1 \over \alpha_\parallel} \left[ \beta_1 \mbox{sin} \psi -u + {\beta_2 d_z \mbox{cos} \psi 
\over d^2_y+d^2_z}\right]^2 
\right)
\]
\begin{equation} \mbox{exp}\left(
- {1 \over \alpha_\perp} \left[
\beta_1 \mbox{cos} \psi 
- { \beta_2 d_z \mbox{sin} \psi \over d^2_y+d^2_z} \right]^2 
-{\beta^2_2 d^2_y \over \alpha_\perp (d^2_y+d^2_z)} +
\Delta \right)
\end{equation}
where
\begin{equation}
\Delta=  {d^2_y \mbox{cos}^2 \psi \over \alpha^2_\parallel \alpha^2_\perp A}
\left[ {\beta_2 d_z \mbox{cos} \psi (\alpha_\perp -\alpha_\parallel) 
\over d^2_y+d^2_z} + \alpha_\perp (\beta_1 \mbox{sin} \psi -u) - \alpha_\parallel \beta_1 \mbox{sin} \psi 
\right]^2
\end{equation}
$d_x$, $d_y$, and $d_z$ come from ${\bf v \cdot n'}= v_x d_x + v_y d_y + v_z d_z$, and are 
given by
\[
d_x = - \mbox{sin} \phi\: \mbox{cos} \theta\: 
\mbox{cos} \psi + \mbox{cos} \phi\: \mbox{sin} \psi,
\]
\[
d_y = -\mbox{sin} \phi\: \mbox{sin} \theta,
\]
\begin{equation}
d_z = -\mbox{cos} \phi\: \mbox{cos} \psi - \mbox{sin} \phi\: \mbox{cos} \theta\: \mbox{sin} \psi
\end{equation}
and
\[
A= {d^2_z \over \alpha_\perp } + d^2_y \left( {\mbox{cos}^2 \psi \over \alpha_\parallel}+ 
{\mbox{sin}^2 \psi \over \alpha_\perp} \right), 
\]
\begin{equation}
\beta_1= {c \over \nu_0} (\nu-\nu_0),~
\beta_2= {c \over \nu_0} \left[ ( \nu_0-\nu') - d_x (\nu_0 - \nu)\right] 
\end{equation}
 
Inspection of (1) and (4) shows that the measured intensity
of the resonantly scattered component, $I_R (\nu)$, depends on the
physical properties of the oxygen ions and the strength
of the chromospheric radiation.
Since the velocity distribution perpendicular to the radial
direction will
have a component along the direction of the incident radiation
from the whole solar disk (except from Sun center), 
$T_\perp$ will also contribute to the resonance scattering.
Consequently, the resonantly scattered component 
of the oxygen lines along the line of sight
results from the combined effects of disk radiation, $T_\perp$
and $T_\parallel$, Doppler dimming and pumping, leading to ratios
similar to those shown in Figure 2.

Figure 2a shows the intensity ratio 1032/1037.6 as a function of
O$^{5+}$ outflow velocity 
obtained by computing the integral along the LOS at 3 $\rm R_s$
of both the resonantly scattered component given in (\ref{eq:rad})
and the collisionally excited component
(see, for example, expression given by Withbroe et al. 1982).
In this example, the electron density is taken to be
4.7$\times 10^4 \mbox{ cm}^{-3}$ at 3 $\rm R_s$,
which is at the upper limit of 
the electron density profile derived from polarization
brightness measurements at that distance (Fisher and Guhathakurta, 1995).
The intensity of the disk radiation
used, $I_{O VI} $ (1032) = 305, $I_{O VI} $ (1037.6) = 152.5, and
$I_{C II} $ (1037) = 52 $erg ~ cm^{-2} ~ s^{-1} ~ sr^{-1}$,
and the 1/e half width of 0.1 and 0.07 {\AA} for the O VI and  C II lines
respectively, were those given by Noci et al. (1987).
The ratio of the C II and O VI line intensities is also consistent
with the more recent values given by Warren et al. (1997).
Contributions to the LOS from larger heliocentric distances
are estimated using that density profile, assuming spherical symmetry,
and a corresponding velocity inferred from the constancy of mass flux.
The electron temperature is kept constant at $10^6$ K.
The parallel and perpendicular temperatures of the O$^{5+}$ ions
are kept constant along the LOS. 
Different combinations of parallel and perpendicular temperatures,
including isotropic and anisotropic temperatures, were
chosen to produce the curves shown in Figure 2a.
The values are within the range expected
from UVCS observations (see Kohl et al., 1997).
The 1032/1037.6 line ratios thus computed include the pumping
of the O VI 1037.6 {\AA} line by C II 1037 {\AA}.
The calculated ratio is found to be insensitive
to the electron temperature or to the elemental abundance.

For an anisotropic kinetic temperature distribution,
Figure 2a clearly shows how the ratio cannot be smaller than 1
if the perpendicular kinetic temperature
is as high as $2.5\times10^8$ K, 
as implied by the UVCS observations (see Kohl et al. 1997). 
At such a high perpendicular kinetic temperature, 
the projection of the perpendicular velocity distribution on the incident 
light rays is strong enough that it 
broadens the spectral width of the C II 1037.018 {\AA} pumping, so
that the pumping by that line alone could only account for a ratio of
1 or larger.
The minimum value of the O VI 1032/1037 ratio in the examples
of anisotropic temperatures in Figure 2a
is $\geq$ 1 even when the parallel temperature is as low as $10^5$ K.

However, UVCS observations in polar coronal holes yield a
ratio of 0.8 at 2.5 $R_s$ and 0.7 at 3 $R_s$
(see Kohl et
al. 1997). Interestingly, observations by SUMER 
(Feldman et al. 1997, Curdt et al. 1997, Warren et al. 1997) 
and UVCS show that there is another 
C II 1036.3 {\AA} line which is almost as strong as the C II 1037 {\AA} 
line. In optically thin conditions, 
the intensity ratio between the C II 1036.3 and 
C II 1037 lines is 0.5 (e.g., Verner et al. 1996). However, 
the C II lines are
emitted from the chromosphere and are probably not optically thin.
Indeed, Sun center disk observations by Warren et al. (1997) yield a ratio of
0.84, while
Curdt et al. (1997) measured a ratio of 0.93 in a field of view
encompassing disk and some coronal emission in a polar coronal hole.
Given that the resonance scattering in the corona is due to the disk
emission, we adopt
the ratio C II 1036.3/1037 of 0.84 given by Warren et al. (1997)
in our calculations.

The additional pumping by this second line is maximum
when the O$^{5+}$ velocity reaches 370 km/s. 
As shown in Figure 2b,
if the pumping of both the C II 1037 and 1036.3 {\AA} lines 
is included, the profile of the
O VI 1032/1037 line ratio becomes very different.
For the examples of isotropic temperatures
(thin solid and dotted lines), the O VI 1032/1037 ratio has
two minima reflecting the pumping by the two C II lines
respectively. However, in the presence of anisotropy
the two mimima merge into a broad one at 370 km/s.
The ratio rises back to a value of 2 at 600 km/s,
as resonance scattering disappears and collisional excitation only remains.

We also note that when pumping by only one C II line is considered,
the different curves in Figure 2a
intersect at a common value of 80 km/s for a ratio of 2.6.
When pumping by both lines is taken into account (Figure 2b),
the curves still intersect at this value,
but also around 400 km/s for a ratio of 1.
Hence, these examples illustrate that the
theoretical value of 2 for the ratio corresponding to 94 km/s
and an isotropic temperature of 1.6 $\times 10^6$ K
(Noci et al. 1987),
is now slightly shifted to a higher ratio and a lower flow speed when the
contribution of the perpendicular temperature
is taken into account in the line of sight integration.

The effect of the pumping by the two C II lines can also be explored
using a self-consistent multifluid solar wind model
(Li et al. 1997) with heating by MHD turbulence (Kraichnan 1965, Li et
al. 1998). 
The results shown in Figure 3 are plotted out to 10 $\rm R_s$
for comparison with UVCS observations, even though the
computations are carried out to 1 AU.
The density of the oxygen ions at the coronal base is chosen such
that the line intensities fit the observed ones there.
The density profile matches that
inferred from polarization brightness measurements in the inner corona
(see Fisher and Guhathakurta, 1995), and
the derived proton velocity of 740 km/s and mass
flux of 1.96 $\times 10^8 \mbox{cm}^{-2}$ at 1 AU
match observational constraints
(e.g. Phillips et al. 1995).
It is seen from Figure 3a,3b that this mechanism can lead to a
significant anisotropy in the inner corona, and that the flow speed of
the oxygen ions exceeds that of protons there. 
The corresponding widths and intensity ratio 1032/1037 are shown in
Figure 3c,3d. Because of the consecutive pumping by the two C II
lines, the resonantly scattered component of  O VI 1037.6 {\AA} will be
enhanced at two discrete values,
leading to a reduction in the width of the 
O VI 1037.6 {\AA} line, which occurs
approximately at 2.25 and 3.25 $\rm R_s$ in this model.
As a result, the O VI 1037.6 {\AA} line will be narrower than
the O VI 1032 {\AA} line.
As shown in Figure 3d, the pumping by the C II 1036.3 {\AA} line
can strongly influence the intensity ratio 1032/1037.6.
When this pumping is neglected (dotted line)
the ratio becomes significantly larger
than 1, but drops below 1, as observed
(see Kohl et al. 1997a, 1997b), when the effect is included.

\section{Conclusions}

This study demonstrates that the anisotropic kinetic temperatures of
O$^{5+}$ implied by UVCS observations (Kohl et al. 1997) in the inner 
corona can significantly affect the Doppler dimming and pumping,
and subsequently
increase the observed O VI 1032/1037.6 line ratio expected for a
given velocity.
Another aspect of this study is that
the pumping by both the C II 1037.018 {\AA} and a second  C II line at
1036.3 {\AA} is necessary to account for the observed intensity
ratios. The
inclusion of this second C II line, however, implies that the O$^{5+}$
ions can reach speeds of 400 km/s near 3 $\rm R_s$.
Consequently, the diagnostic techniques used in UVCS observations can
be extended to much higher flow speeds than originally expected.

We also note that protons, which are strongly coupled to the
neutral hydrogen, are expected to have the same flow speed as the neutrals (see
Olsen et al. 1994, Allen et al. 1998). Hence, based on the recent UVCS
observations, this study, together with the empirical model results
by Kohl et al. (1998), shows that the O$^{+5}$ ions
are flowing significantly faster than the protons around 
3 $R_s$ in the fast solar wind. 
This is an indication that whatever mechanism is responsible for
coronal heating and solar wind acceleration must preferentially 
accelerate minor ions such that they can
flow faster than protons in the inner corona.

\acknowledgements 

The results presented in this Letter were motivated by the recent UVCS
observations. We extend our thanks to the UVCS team for
the outstanding value of their measurements.
We gratefully acknowledge helpful discussions with Drs.
N. S. Brickhouse, S. Cranmer, 
J. C. Raymond, and L. Strachan, and thank M. Romoli for
pointing out to us a correction to Equation (1) as it appeared
in an earlier version of the manuscript.
UVCS is supported by the National Aeronautics and Space Administration
under grant NAG5-3192 to the Smithsonian Astrophysical Observatory, by
Agenzia Spaziale Italiana, and by Swiss funding agencies.
X. Li and S. R. Habbal were supported by NASA grants NA65-6271 and NA65-6215 
to the Smithsonian Astrophysical Observatory. 
X. Li was also supported by a predoctoral 
fellowship from the Smithsonian Astrophysical Observatory. 

\section{References}

\parindent=0.in

Allen, L., Habbal, S.R., and Hu, Y.-Q. 1998, JGR, in press

Curdt, W., et al., 1997 A\&AS, 126, 281

Feldman, U., et al. 1997, ApJS, 113, 195

Fisher, R.R. and Guhathakurta, M. 1995, ApJ, 447, L139

Habbal, S.R., et al. 1997, ApJ, 489, L103 

Hyder, C.L. \& Lites, B.W. 1970, Sol. Phys., 14, 147

Kohl, J.L. \& Withbroe, G.L. 1982, ApJ, 256, 263

Kohl, J.L. et al. 1995, Sol. Phys., 162, 313

Kohl, J.L. et al. 1997, Sol. Phys., 175, 613

Kohl, J. L. et al. 1998, ApJ (this issue)

Kraichnan, R. 1965, Phys. Fluids, 8, 1385

Li, X., Esser, R., Habbal, S.R. \& Hu, Y.-Q. 1997, J. Geophys. Res., 
102, 17419

Li, X., Habbal, S.R., Esser, R., \& Hollweg, J. V. 1998 (in preparation)

Noci, G., Kohl, J.L. \& Withbroe, G.L. 1987, ApJ, 315, 706

Noci, G. et al. 1997, ASP Conference Series, in press

Olsen, E.L., Leer, E., Holzer, T.E. 1994, ApJ, 420, 913

Phillips, J. L., et al., 1995, Science, 268, 1030

Verner, D.A., Verner, E.M., Ferland, G.J. 1996, Atomic Data and Nuclear
Data Tables, 61, 1

Warren, H.P., Mariska, J.T., Wilhelm, K., and Lemaire, P. 
1997, ApJ, 484, L91

Withbroe, G.L., Kohl, J.L., Weiser, H., \& Munro, R.H. 1982,
Space Sci. Rev., 33, 17

\begin{figure}
\epsscale{0.7}
\plotone{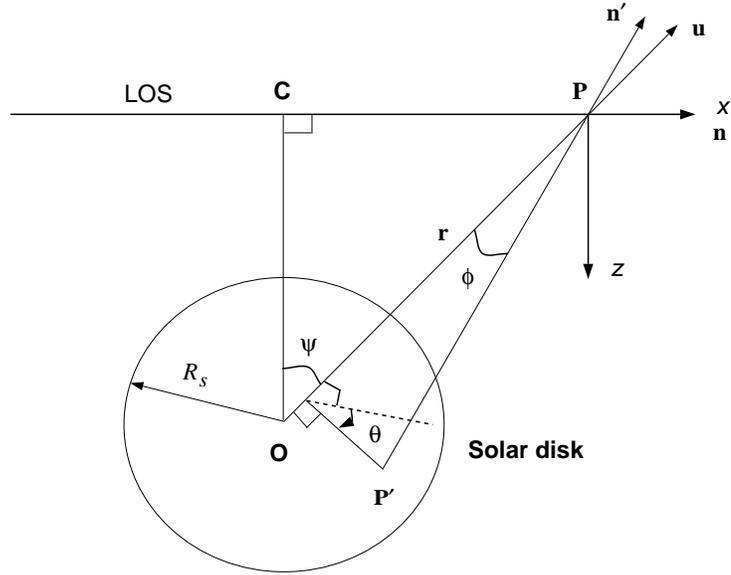}
\caption{Coordinate system used to derive Equation (5).
LOS is the line of sight chosen along the x-axis.
C represents the point of closest approach from the LOS to the Sun.
P is a point along the LOS which contributes to the integral in (1).
$T_{\parallel}$ is the component of the kinetic temperature along the radial
direction {\bf r}.
{\bf r} is also the direction of the outflow velocity {\bf u}.
$T_\perp$ is the component of the kinetic temperature in a plane
perpendicular to {\bf r} at P.
P$'$ is a point on the solar disk from which the radiation is emitted.
$\theta$ is the angle between the x-z plane and the plane OPP$'$.
The projection of P$'$ onto that plane is shown by the dashed line.
The direction of the
radiation from P$'$ reaching P is indicated by the unit vector {\bf n$'$}.
\label{Fig. 1}}
\end{figure} 

\begin{figure} 
\epsscale{0.9}
\plotone{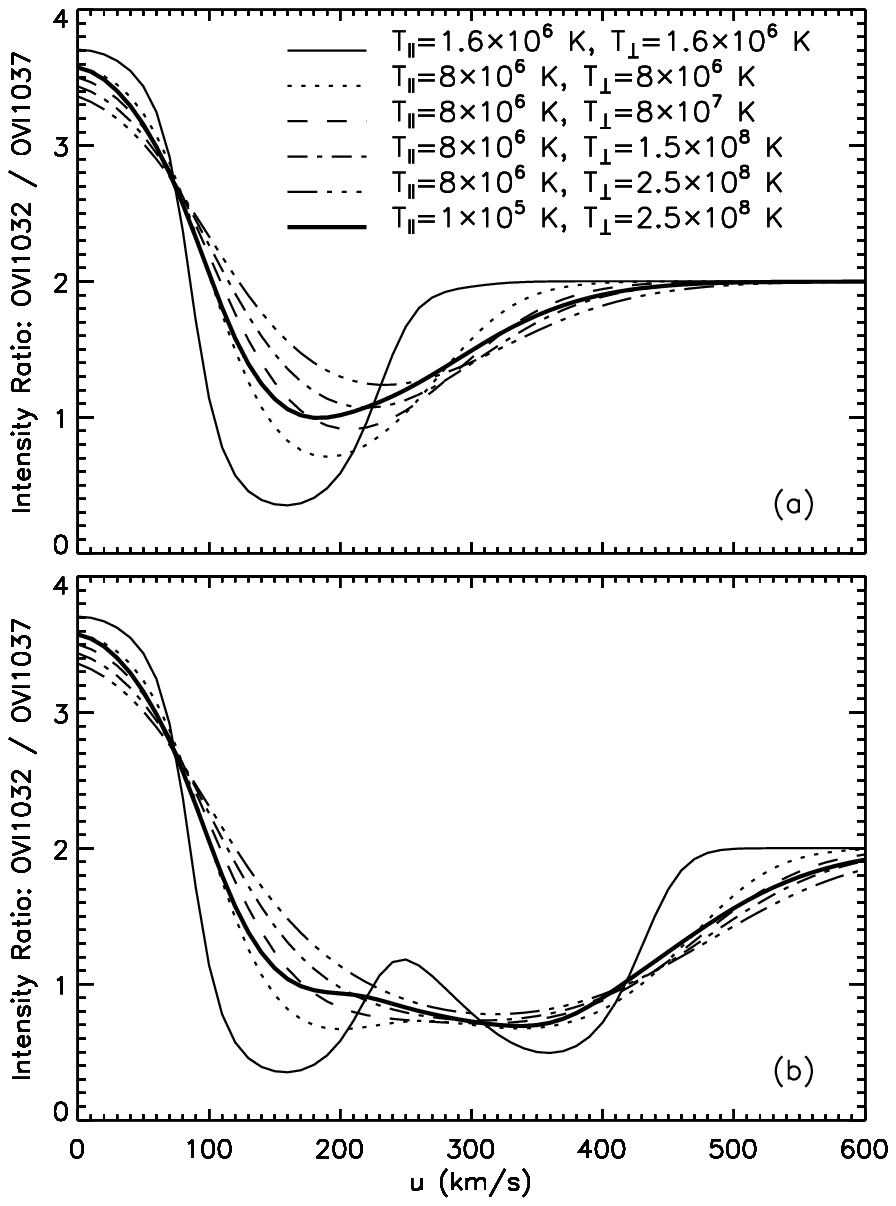}
\caption{Computed O VI 1032/1037.6 line ratio at r = $3 ~ R_s$ as a function of
outflow velocity, for different combinations of
$T_{\parallel}$ and $T_\perp$, including pumping
of the O VI 1037.6 {\AA} line by C II 1037 {\AA} only (a) or by both
C II 1037 {\AA} and 1036.3 {\AA} lines (b).
The electron density is 4.7 $10^4 ~ cm^{-3}$ at $3 ~ R_s$.
\label{Fig. 2}}
\end{figure}

\begin{figure}
\plotone{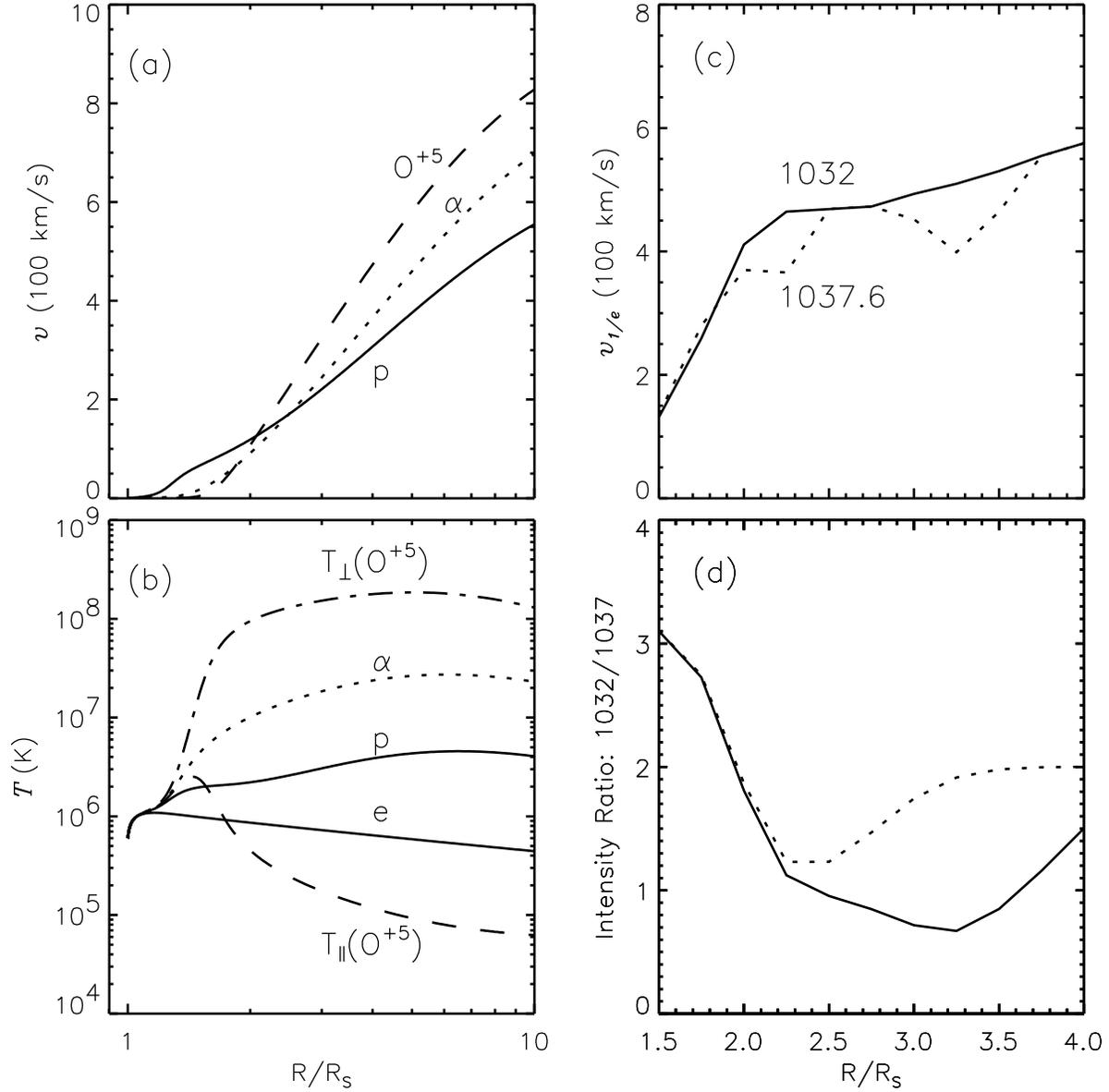}
\caption{Multi-fluid solar wind model computations
for electron/protons, alpha particles and oxygen ions:
(a) flow speeds,
(b) temperature of species, (c) line
widths, $v_{1/e}$, and (d) intensity ratio.
In (d) the dotted line corresponds to
pumping by only C II 1037.0 {\AA}, while the solid line includes the
pumping by both  C II 1037.0 and 1036.3 {\AA} lines. 
%$Q_{e0} ~ = ~ 1.2 ~ \times ~ 10^{-13} ~ \mbox{erg cm}^{-3}
%\mbox{s}^{-1},~\lambda_e ~ = ~ 0.2 ~ \rm R_s$, and  
%$\Gamma_p ~ = ~ 31.375$
\label{Fig. 3}}
\end{figure}

\end{document}